\documentclass[iop]{emulateapj} 

\usepackage{natbib}
\usepackage{graphicx}
\usepackage{amsmath}
\usepackage{color}
\usepackage{ulem}
\graphicspath{{./figs/}}

\newcommand{\Msun}{{\rm M_{\odot}}}

\newcommand{\Zsun}{Z_{\odot}}

\newcommand{\Mbh}{M_{\rm BH}}

\newcommand{\ad}{a_{\rm d}}

\newcommand{\td}{T_{\rm d}}

\newcommand{\rhii}{r_{\rm HII}}
\newcommand{\feddt}{\left< f_{\rm Edd} \right>}
\newcommand{\fedd}{f_{\rm Edd}}
\newcommand{\cs}{c_{\rm s}}
\newcommand{\Ledd}{L_{\rm Edd}}
\newcommand{\fdust}{f_{\rm d}}

\newcommand{\frade}{f_{\rm rad}^{\rm e}}
\newcommand{\fradd}{f_{\rm rad}^{\rm dust}}
\newcommand{\fradhi}{f_{\rm rad}^{\rm HI}}
\newcommand{\fgrav}{f_{\rm grav}}
\newcommand{\xhi}{x_{\rm HI}}

\newcommand{\tdust}{T_{\rm dust}}
\newcommand{\rb}{r_{\rm B}}
\newcommand{\nh}{n_{\rm H}}
\newcommand{\nhinf}{n_{\rm \infty}}

\begin{document}   

%
%
\title{Dusty gas accretion onto massive black holes and infrared diagnosis of the Eddington ratio}
%

%
%
\author
{
Hidenobu Yajima\altaffilmark{1,2},
Massimo Ricotti\altaffilmark{3},
KwangHo Park\altaffilmark{4},
Kazuyuki Sugimura\altaffilmark{2},
}

\affil{$^{1}$Frontier Research Institute for Interdisciplinary Sciences, Tohoku University, Sendai, Miyagi 980-8578, Japan}

\affil{$^{2}$Astronomical Institute, Tohoku University, Sendai, Miyagi 980-8578, Japan}

\affil{$^{3}$Department of Astronomy, University of Maryland, College Park, MD 20740, USA}

\affil{$^{4}$Center for Relativistic Astrophysics, School of Physics, Georgia Institute of Technology, Atlanta, GA 30332, USA}

\email{yajima@astr.tohoku.ac.jp} 

 

\label{firstpage}
%
%
\begin{abstract}
Evidence for dust around supermassive black holes (SMBHs) in the early Universe is strongly suggested by recent observations.
However, the accretion mechanism of SMBHs in dusty gas is not well understood yet.
We investigate the growth of intermediate-mass black-holes (IMBHs) of $\sim 10^{5}~\Msun$ in dusty clouds by using one-dimensional radiative-hydrodynamics simulations.
We find that the accretion of dusty gas onto IMBHs proceeds gently with small fluctuations of the accretion rate, whereas that of pristine gas causes more violent periodic bursts.
At dust-to-gas mass ratios similar to the solar neighborhood, the time averaged luminosity becomes smaller than that for primordial gas by one order of magnitude and the time-averaged Eddington ratio ranges from $\sim 10^{-4}$ to $\sim 10^{-2}$ in clouds with initial gas densities of $\nh = 10 - 1000~\rm cm^{-3}$.
Our calculations show that the effect of dust opacity alone is secondary compared to the radiation pressure on dust in regulating the BH growth. 
We also derive spectral energy distributions at IR bands by calculating dust thermal emission and show that the flux ratio between $\lambda \lesssim 20~\rm \mu m$ and $\gtrsim 100~\rm \mu m$ is closely related to the Eddington ratio. 
Thermal emission from hot dust near the BH dominates only during the high accretion phase, producing higher flux density at $\lesssim 20~\rm \mu m$. 
Therefore, we suggest that the combinations of MIR observations by JWST and FIR observation by ALMA or Spitzer can be used
to estimate the Eddington ratio of massive BHs.
\end{abstract}

%
%

%
%
\section{Introduction}
Understanding the growth mechanism of massive black holes is one of
the major challenges of modern astrophysics.  It is known that the number
density of supermassive black holes (SMBHs) steeply decreases in the
early Universe \citep[e.g.,][]{Richards06}.  Yet, recent observations
have detected SMBHs with masses $\gtrsim 10^{9}~\Msun$ even at cosmic
times less than $\rm \sim 1~Gyr$ \citep{Mortlock11, Wu15}. 
However, the existence of SMBHs at high redshifts alone does not suggest clues 
about the initial seed mass and the subsequent growth history. 

Various mechanisms have been suggested for the formation
channel of SMBH seeds, that typically are assumed to have masses
$\Mbh \sim 10^{5}~\Msun$: (1) growth from stellar mass BHs of
Population III star remnants by gas accretion \citep{Alvarez09,
  Jeon12, Park13}; (2) direct collapse of supermassive stars
\citep{Omukai01, Begelman06, Volonteri10, Agarwal12, Latif13,
  Inayoshi14a, Sugimura14}; (3) formation in dense star clusters via
collisions among stars \citep{Rees78, Portegies-Zwart02, Devecchi12,
  Katz15, Yajima16a}.  After the formation of SMBH seeds, it is widely
believed that further growth proceeds though gas accretion.  However,
the accretion mechanism has not been understood yet because the
accreting gas is subject to feedback from radiation emitted near the
accretion disks around BHs.

One of the mechanisms limiting the gas accretion rate is the balance
between gravitational attraction and radiation pressure on
free-electron, the so-called Eddington limit.  Cosmological
simulations showed that massive black holes of $\sim 10^{5}~\Msun$
could grow up to supermassive ones of $\sim 10^{9}~\Msun$ by $z \sim
6$ using Eddington limited Bondi-Hoyle accretion prescription and a simple
thermal feedback model \citep{Li07, DiMatteo08, DiMatteo12,
  Sijacki09}.  \citet{DiMatteo12} carried out cosmological simulations
in large cosmological volumes of $(0.75~\rm Gpc)^{3}$ in comoving unit
and showed the accretion rate of SMBHs in rare massive galaxies was near
the Eddington limit most of the time, resulting in SMBHs with $\sim
10^{9}~\Msun$ at $z \sim 7$.  However, due to computational
limitations, in such cosmological simulations gas dynamics at the
Bondi radius (where the gravity of BHs is dominant) is not well
resolved, thus requiring sub-grid feedback models.  Therefore there is
a large uncertainty in the estimation of the gas accretion rate.

Using high-resolution radiation hydrodynamics simulations resolving
the Bondi radius, \citet{Milosavljevic09a} showed that 
the neighboring gas is ionized by the radiation from a central BH
and the thermal pressure of H{\sc ii} regions pushed gas
away from the BH against the gravity.  As a result, the accretion rate
was significantly suppressed even at lower luminosities than
Eddington.  \citet{Park11} showed that the gas accretion periodically
changed and the time-averaged accretion rate was $\sim 1~\%$ of the
Bondi rate regardless of some parameters, e.g., radiative efficiency,
black hole mass, background density \citep[see also,][]{Park12}.

 \citet{Inayoshi16} suggested that gas accretion rate could exceed the
 Eddington limit when BHs accrete from extremely high-density gas
 clouds where the size of ionized bubble is smaller than Bondi radius
 \citep[see also,][]{Park14a, Sakurai16}.  In addition, assuming the
 anisotropic radiation feedback, \citet{Sugimura16} showed that gas
 efficiently accretes onto a BH along the shadowed region and the
 accretion rate exceeds the Eddington limit.  Thus, it is still
 unclear how much the growth of BHs is regulated by radiative
 feedback.

In local galaxies, it is well known that BH mass tightly correlates
with bulge mass or velocity dispersion \citep{Kormendy13}.  This
implies co-evolution of BHs with galaxies. \cite{Park16} showed that
the growth rate of massive BHs can be enhanced under the influence of
gravitational potential of the bulge.  In addition, as star formation
proceeds, gas surrounding a BH is metal/dust enriched through
type-I/II supernovae, and stellar winds.  Observations of
high-redshift quasars at $z \gtrsim 6$ suggest that a large mass in dust exists
around SMBHs. Dust is detected via its thermal emission or dust
extinction \citep{Bertoldi03, Priddey03, Maiolino04, Wang13}.
In addition, recent discoveries of high-redshift sub-millimeter
galaxies indicate that some galaxies can become dust rich at early times \citep{Riechers13, Watson15}.  
Theoretically, recent simulations show that the metallicity near the
galactic centers could reach $\sim 0.01~\Zsun$ even at $z \sim 10$
\citep[e.g.,][]{Wise12a} or even higher depending on the halo mass \citep[e.g.,][]{Ricotti05, Ricotti16}. 

The metallicity and dust amount of massive galaxies in an over-dense
region could reach the level of solar-neighborhood even at $z \gtrsim
6$ \citep[e.g.,][]{Yajima15c}.  Therefore BHs were likely to grow in a
dusty medium even in the early Universe.  If dust exists around a BH,
radiation from the inner parts of an accretion disk around a BH can be
obscured.  This changes the observational properties of accreting BHs
and the dynamics of accreting gas.  In addition to dust opacity, the
radiation force on dust can play a roll in determining the growth rate
of BHs \citep{Ciotti07, Namekata14, Hensley14}.  However, the
interplay between dust and the photo-ionization feedback, which is the
main feedback mechanism suppressing the growth of stellar or
intermediate-mass BHs (IMBHs), has not been studied in
sufficient detail.  In this work we investigate the impacts of dust on
the growth of BHs by using one-dimensional radiation hydrodynamics
simulations resolving the both the Bondi radius and ionized bubbles
simultaneously. We also estimate self-consistently the thermal
emission from dust in our time-dependent models and show that the
emission from a hot dust component at $\sim 20 \rm \mu m$ is prominent
only during luminosity bursts, while a warmer dust component produces
a flux at $100 \rm \mu m$ that is roughly proportional to the mean accretion rate.
We thus conclude that the flux ratio at $20 \rm \mu m/100 \rm \mu m$ is a good
proxy for the Eddington ratio.

The paper is organized as follows. We describe our models in \S2.  In
\S3, we present simulation results that include time-average Eddington
ratios, accretion histories with and without dust, dependences of the
Eddington ratios on metallicity and BH mass, spectral energy
distributions (SEDs) in the infrared (IR) considering dust thermal
emission.  We discuss the dust destruction processes and the condition
for hyper-accretion in \S5, and summarize our results in \S6.

%
%
\section{Model}
\label{sec:model}
We solve the dynamics of gas surrounding a BH under radiative feedback
using one-dimensional radiation hydrodynamics simulations. 
In this work, we use a hydrodynamics code Zeus-MP \citep{Stone92, Hayes06}. 
\citet{Park11} incorporated the radiative transfer of X-ray and UV photons 
and chemical reactions of primordial gas into the Zeus-MP. 
Here we furthermore incorporate dust attenuation and radiation pressure on dust to the code. 
In spherical symmetric coordinates the basic equations, i.e., the conservations of mass, momentum and energy, are:
\begin{equation}
\begin{split}
&\frac{\partial \rho}{\partial t} + \frac{1}{r^{2}} \frac{\partial}{\partial r} (r^{2} \rho v) = 0, \\
&\rho \left( \frac{\partial v}{\partial t}  + v \frac{\partial v}{\partial r} \right)
= - \frac{\partial p}{\partial r}  - \frac{G\Mbh \rho}{r^{2}} + f_{\rm rad}, \\
&\rho \left(  \frac{\partial e}{\partial t} + v \frac{\partial e}{\partial r} \right)
= - p \frac{1}{r^{2}} \frac{\partial}{\partial r} \left( r^{2} v \right) + \Gamma - \Lambda,
\end{split}
\end{equation}
where $f_{\rm rad}$ is radiation force,
$\Gamma$ is the heating rate by photo-ionization of hydrogen and helium, and $\Lambda$ is the radiative cooling rate. 
In this work, we consider radiation force on dust ($\fradd$), free electrons ($\frade$), and neutral hydrogen ($\fradhi$).
For single-sized grains of dust the optical depth is
\begin{equation}
\begin{split}
d\tau_{{\rm d}, \nu} = Q_{\rm \nu} \pi a_{\rm d}^{2} \frac{D m_{\rm H} \nh}{m_{\rm d}} dl
= \frac{3 Q_{\rm \nu} D m_{\rm H} \nh}{4\rho_{\rm d} a_{\rm d}} dl,
\end{split}
\label{eq:taudust}
\end{equation}
where $a_{\rm d}$ is dust radius, $m_{\rm d}$ is mass of a dust grain,
$Q_{\rm \nu}$ is the absorption coefficient to geometrical cross
section, $m_{\rm H}$ is hydrogen mass, $\rho_{\rm d}$ is mass density
of a dust grain, and $D$ is dust-to-gas mass ratio. If the wavelength
of the radiation is shorter than $\sim 2 \pi a_{\rm d}$, then $Q_{\rm
  \nu} \sim 1$ \citep{Draine84}.  Here we consider compact spherical
dust grains, i.e., $m_{\rm d} = 4 \pi a_{\rm d}^{3} \rho_{\rm d} / 3$.
Here we assume that the dust-to-gas mass ratio is proportional to
metallicity as $D \equiv M_{\rm dust}/M_{\rm gas} = 0.01 \left(
{Z}/{\Zsun} \right)$, motivated by observations of a nearly
constant dust-to-metal mass ratio in local galaxies \citep{Draine07}.

Observations of local galaxies indicate that dust has a continuum size distributions. For example, the dust in the Milky Way shows a power-law size distribution ${dn_{\rm d}}/{da_{\rm d}} \propto a_{\rm d}^{-3.5}$, often referred to as the MRN distribution \citep{Mathis77}. However, for sake of simplicity here we use a single dust size model with a fiducial size $a_{\rm d} = 0.1~\rm \mu m$. The choice of the fiducial size is motivated below.
At wavelengths $\lambda< 2 \pi a_{\rm d}$, where $Q_{\rm \nu}=1$, the opacity of a power law distribution of dust radii ${dn}/{da} \propto a^{-\alpha}$, is the same of a single size dust model for 
\begin{equation}
a_{\rm d, 1} = \left(  \frac{3 - \alpha}{4 - \alpha} \right)
\left( \frac{a_{\rm d,max}^{4-\alpha} - a_{\rm d,min}^{4-\alpha}}
{a_{\rm d,max}^{3-\alpha} - a_{\rm d, min}^{3-\alpha}} \right),
\end{equation}
where $a_{\rm d, 1}$ is the dust size of the equivalent single-size
model, $a_{\rm d}^{\rm min}$ and $a_{\rm d}^{\rm max}$ are minimum and
maximum grain sizes in the grain distribution.  Assuming realistic
minimum and maximum dust sizes of $a_{\rm d}^{\rm min}=8.1\times10^{-3}~\rm \mu m$ and
$a_{\rm d}^{\rm max}=1.0~\rm \mu m$ in a MRN model ({ i.e.},
$\alpha=3.5$), the equivalent single dust size model has $a_{\rm d,1} =
0.1~\rm \mu m$.  

Near active galactic nuclei (AGNs), the main components of dust are graphite
and silicate because of their high sublimation temperature $T \gtrsim
1500~\rm K$, whereas icy dust can be easily sublimated.  Here we
consider a mix of graphite and silicate dust grains with the mass
ratio of $1:1$. The difference of mass density between the graphite and silicate is small. In addition, when $Q_{\nu}=1$, the absorption cross section depends only on the surface area of a dust grain. Therefore, the simulation results are rather insensitive to the assumed mass ratio of the dust components.

Within fully ionized region, the radiation force on dust can be larger than that from Compton scattering on electrons by a factor
\begin{equation}
\begin{split}
f_{\rm d} \equiv \frac{\fradd}{\frade} &= \frac{3Q_{\rm \nu}Dm_{\rm H} }{4 \rho_{\rm d} a_{\rm d} \sigma_{\rm T}} \\
&=7.1 \times 10^{2} 
\left( \frac{\ad}{0.1~\rm \mu m} \right)^{-1}
\left( \frac{Z}{\Zsun} \right).
\end{split}
\label{eq:fratio}
\end{equation}
UV photons absorbed by dust are reemitted as IR photons via thermal
emission of dust.  However, the IR photons can escape from neighbor
gas without further interaction with dust due to the lower absorption
cross-section to the IR wavelengths.  
If the hydrogen column density exceeds $10^{22}~\rm cm^{-2}$, the IR photons can be absorbed by dust
and impart additional momentum to the gas. In this work, we focus on spatial scales up to an ionized bubble. On these scales the absorption of IR photons by dust is negligible. Therefore, we neglect dust opacity to IR light.

Aside from the addition of dust physics, the simulations have the same physics and initial conditions
as the ones presented in \citet{Park12}.  For our fiducial simulations
we investigate the gas dynamics around BHs of $10^{5}~\Msun$ embedded
in gas clouds with uniform initial gas density.  The radiative
luminosity is  $L=\eta \dot{m}c^{2}$ with a constant
radiative efficiency $\eta=0.1$ and $\dot m$ estimated at the inner
boundary in the logarithmically spaced radial grid.  We assume a
power-law spectrum $L_{\rm \nu} \propto \nu^{-1.5}$ with the frequency
range $\nu \ge \nu_{\rm L}$, where $\nu_{\rm L}$ is the Lyman-limit
frequency. However the results can be easily scaled to different mass
BHs and different densities as the problem is basically scale-free.
\citet{Park12} showed that simulations with a fixed $\Mbh \nh$ show
identical behavior once the length and time scales are re-normalized
appropriately.  \citet{Park12} used the condition of $\Mbh \nh =
10^{7}~\rm \Msun~cm^{-3}$ as a fiducial run.  For this parameters combination, if
gas accretes onto the BH at the Bondi rate, the luminosity is close
to the Eddington luminosity.  Here we use the same fiducial run with
$\Mbh \nh = 10^{7}~\rm \Msun~cm^{-3}$ to compare the impact of dust
attenuation and radiation force on the gas accretion onto BHs to the dust-free case.
Table~\ref{table:model} summarizes three different models we ran to
understand the impact of dust.: (1) dust-free case (M5-Z0), (2) dusty
case with $Z=\Zsun$ without radiation pressure on dust (M5-Z1), and (3)
dusty case with $Z=\Zsun$ including radiation pressure on dust
(M5-Z1rad).  In all simulations we include $\frade$ and $\fradhi$.  The
inner and outer boundaries are
$2\times10^{-2}$~pc and $2\times10^{3} ~ \rm pc$, respectively.  
The number of cells is $n_{\rm cell} = 256$. 
We have checked that our results are convergent and do not change when increasing the number of cells from $128$ to $400$. 
The cell size changes with radial
distance with the equatorial ratio.  To check that indeed the results
are scalable to different BH masses we run simulations with BH masses
of $10^{3}, 10^{4}$ and $10^{6}~\Msun$ keeping constant $\Mbh
\nh=10^{7}~\rm \Msun~cm^{-3}$.

\begin{table}
\begin{center}
 \caption{Model parameters}
\begin{tabular}{ccccccc}
\hline
Model &  $\Mbh~[\Msun]$ & $\nhinf~[\rm cm^{-3}]$ &Dust attenuation & $\fradd$ \\
\hline
M5-Z0      & $10^{5}$ & 100 & $\times$ & $\times$\\
M5-Z1     & $10^{5}$ & 100 & $\bigcirc$ & $\times$ \\
M5-Z1rad     & $10^{5}$ & 100  & $\bigcirc$ & $\bigcirc$ \\
\hline
\end{tabular}
\begin{flushleft}
NOTES. $\nhinf$ is initial background density. $\fradd$ is radiation pressure on dust.
\end{flushleft}
\label{table:model}
\end{center}
\end{table}


%
%

\section{Results}
\label{sec:result}

\subsection{Time evolution of luminosity under radiative feedback}
\label{sec:mdot}

\begin{figure}
\begin{center}
\includegraphics[scale=0.4]{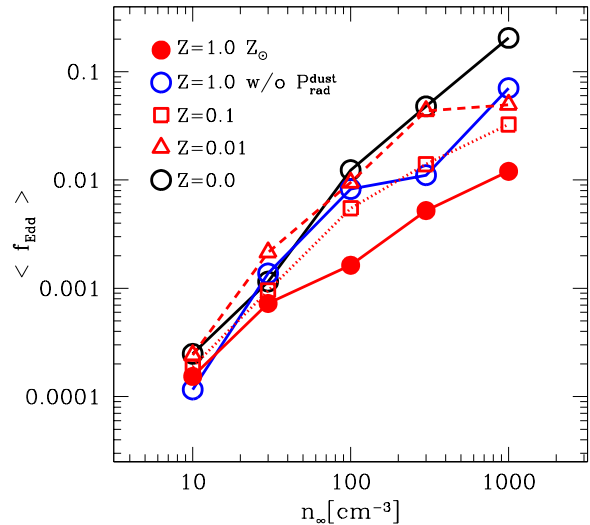}
\caption{
Time-averaged luminosities in Eddington units, $\feddt$, as a function of initial density of the ambient medium for a $10^5$~M$_\odot$ BH. 
Black open circles show $\feddt$ in the case without dust.
Red symbols refer to simulations with dust and metallicities $Z=0.01~\Zsun$ (open triangles),
$0.1~\Zsun$ (open squares) and $1.0~\Zsun$ (filled circles). 
Blue open circles refer to $\feddt$ in dusty gas including dust opacity but without radiation force on dust. 
}
\label{fig:mdot}
\end{center}
\end{figure}
 Figure~\ref{fig:mdot} shows time-averaged BH luminosity $\feddt
 \equiv {\left< L \right>}/{\Ledd}$, in units of the Eddington luminosity $\Ledd
 \cong 1.3\times10^{43}~{\rm erg~s^{-1}}~\left( {\Mbh}/{10^{5}~\Msun}
 \right)$, for the simulations in Table~\ref{table:model}.  We
 calculate $\feddt$ at various metallicities ($Z=0 - 1.0~\Zsun$), and
 initial gas densities $\nhinf = 10 - 1000~\rm cm^{-3}$.  Regardless
 of dust amount, $\feddt$ linearly increases with gas density.  This
 means that the accretion rate normalized by the Bondi rate is
 constant as a function of the gas density.  The Eddington ratio
 $\feddt$ for gas of primordial composition and $\nhinf = 100~\rm
 cm^{-3}$ ({i.e.}, M5-Z0 run), is $ \sim 1\%$ of the Bondi rate.
Thus for the case with $Z=0~\Zsun$, if $\feddt <1$ we have:
\begin{equation}
\feddt(Z=0) \approx 1\% \left( \frac{\Mbh\nh }{10^{7}~\Msun~\rm cm^{-3}} \right).
\end{equation}
This value is consistent with previous studies
 \citep{Milosavljevic09a, Park11, Park12}.  As the metallicity and
 dust content increases, $\feddt$ decreases.  At $Z=10^{-2}~\Zsun$,
 $\feddt$ is similar to the cases without dust.  When the metallicity
 is higher than $\sim 0.1~\Zsun$, $\feddt$ becomes smaller than that
 of primordial gas by a factor greater than $\sim 2$.  The M5-Z1rad run
 with radiation pressure on dust and solar metallicity has $\feddt
 \sim 10^{-3}$, while the same model without radiation pressure on
 dust (M5-Z1) has $\feddt \sim 7 \times 10^{-3}$ that is similar to the
 M5-Z0 run without dust.  This suggests that the radiation force on dust
 significantly reduces the accretion rate, whereas the effect of dust
 opacity alone is secondary.

\begin{figure}
\begin{center}
\includegraphics[scale=0.4]{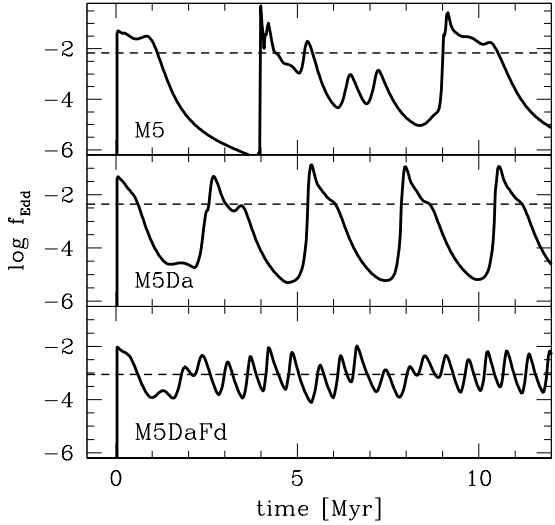}
\caption{ Time evolution of Eddington ratio $\fedd$ for the models in
  Table~\ref{table:model}.  The top, middle, and lower panels show
  $\fedd$ for models M5-Z0, M5-Z1, and M5-Z1rad.  The Dashed lines
  are the time averaged Eddington ratios shown in
  Fig.~\ref{fig:mdot}.}
\label{fig:acc}
\end{center}
\end{figure}

Figure~\ref{fig:acc} shows the time evolution of the luminosity for
the simulations in Table~\ref{table:model}.  As shown in previous
works \citep{Park11}, the accretion of primordial gas periodically
changes.  During the bursts the luminosity is near 
the Eddington limit.  When the luminosity is high, the H{\sc ii}
region expands far beyond the Bondi radius.  At the ionizing front,
the low-density hot ionized gas is almost in pressure equilibrium with
the outside H{\sc i} gas, and suppress the gas inflow from the neutral
region.  This leads to a decrease of accretion rate.  The minimum
luminosity becomes smaller than the maximum one by about 5-6 orders of
magnitude.

 After each burst, the gas density within the H{\sc ii} bubble
  decreases with time.  In addition, the lowered luminosity allows
recombination of hydrogen in H{\sc ii} region with radiative cooling.
Due to the recombination, the ionization degree and temperature in the
H{\sc ii} region decrease.  These break the pressure equilibrium
and allows gas inflow, resulting in the burst of luminosity again.

The mid panel shows the luminosity of M5-Z1 run (with dust).  The burst
cycle is shorter than that of M5-Z0 run (without dust).  This is because
the cycle time scale is proportional to the sound crossing time over
ionized region $\sim \rhii / c_{\rm s}$ \citep{Park12}, and the size
of H{\sc ii} region ($\rhii$) is decreased due to the dust
attenuation.  In order for the dust attenuation to work, the optical
depth of dust in a H{\sc ii} region should be higher than unity.  We
can roughly estimate the critical metallicity (and therefore dust
abundance) that can have an effect at reducing the size of H{\sc ii}
region and therefore the period between bursts.  First we estimate the size of Str\"{o}mgren sphere:
\begin{equation}
\begin{split}
r_{\rm HII} &= \left(  
\frac{3 \fedd \Ledd} {4 \pi \nh^{2} \alpha_{\rm B}\bar{e}_{\rm ion}} 
\right)^{\frac{1}{3}} \\
&=33~{\rm pc}~ \left( \frac{\fedd}{10^{-2}} \right)^{\frac{1}{3}}
\left( \frac{\nh}{10^{2}~\rm cm^{-3}} \right)^{ - \frac{2}{3}}
\left( \frac{\Mbh}{10^{5}~\Msun} \right)^{\frac{1}{3}}, 
\label{eq:rhii}
\end{split}
\end{equation}
where $\alpha_{\rm B}$ is the case-B recombination coefficient, and
$\bar{e}_{\rm ion}$ is the mean energy of ionizing photons. We here
set $T=7\times10^{4}~\rm K$ as the temperature of ionized region in
the above estimation.  For a power-law spectrum, $\bar{e}_{\rm ion}$
is estimated by $\frac{\alpha}{\alpha-1} \times13.6~{\rm eV}$, where
$\alpha$ is a slope of the power-law spectrum, and it is $40.8~\rm eV$
for $\alpha=1.5$.  Using the estimated size of ionized region above,
we can derive the critical metallicity for the dust attenuation as
\begin{equation}
\tau = \frac{3 \times 10^{-2} Q_{\rm \nu} m_{\rm H} \nh}{4\rho_{\rm
    d} a_{\rm d}} \left( \frac{Z_{\rm crit}}{\Zsun}\right) \rhii \approx 1.
\end{equation}
Thus, the critical metallicity is given by
\begin{equation}
\begin{split}
Z_{\rm crit} 
\approx 0.2~{\Zsun}~ 
\left( \frac{\fedd}{10^{-2}} \right)^{-\frac{1}{3}}
\left( \frac{\Mbh\nh }{10^{7}~\Msun~\rm cm^{-3}} \right)^{ - \frac{1}{3}}\\
\approx 0.2~{\Zsun}~\left( \frac{\Mbh\nh }{10^{7}~\Msun~\rm cm^{-3}} \right)^{ - \frac{2}{3}}.
\end{split}
\label{eq:zcrit}
\end{equation}
If the metallicity is higher than $Z_{\rm crit}$, the size of H{\sc
  ii} region decreases and so does the period between bursts.  Note
that, even in the case with the dust attenuation (but no radiation
pressure on dust), the size of ionized bubble is larger than the Bondi
radius.  Therefore, the gas density inside the H{\sc ii} region is
roughly the same as the run with no dust (M5-Z0), and so is the mean
accretion rate.  However, the peak luminosities are somewhat smaller
than in the M5-Z0 run due to the smaller H{\sc ii} bubbles.  This might
lead to the lower $\feddt$ than the primordial gas cases by a factor
$\sim 2$.

The lower panel shows the luminosity for the run that includes the
effect of both dust opacity and radiation pressure on dust (M5-Z1rad).
Unlike M5-Z0 and M5-Z1 runs, the gas accretion of M5-Z1rad proceeds more
gently and the luminosity varies between the burst and quiescent
phases by one or two orders of magnitude (instead of 5 orders of
magnitude as in the other runs).  The maximum and minimum luminosities
are $\sim 10^{-2}$ and $10^{-4}$ of $\Ledd$, respectively.  Due to the
high radiation force on dust that, as shown in Eq.~(\ref{eq:fratio}),
at solar metallicity is nearly 700 times larger than Compton
scattering on electrons, the gas inflow is significantly suppressed
when $\fedd \gtrsim 10^{-3}$, resulting in the lower maximum
luminosity than in the M5-Z0 run by one order of magnitude.  In the case
of M5-Z1rad run, the period between bursts is not determined by the
sound crossing time of the H{\sc ii} region ($\rhii / \cs$).  As will
be shown in Sec.~\ref{sec:rhii}, the H{\sc ii} bubble in the M5-Z1rad
run maintains almost constant size, whereas the one in the M5-Z0 run
recombines and collapses when the luminosity drops to the minimum
\citep[see also,][]{Park12}.  In the case of primordial gas, the
expansion of the H{\sc ii } bubbles significantly reduces the gas
density and accretion rate, and then the rapid gas inflow during the
collapse of the ionized bubble allows the luminosity to reach nearly
the Eddington limit.  On the other hand, in the case of dusty gas, the
dust attenuation regulates the expansion of the H{\sc ii} bubble, and
the radiation force suppresses the gas inflow significantly.  Thus,
the dust plays a role in diminishing the large periodic variability of
the luminosity and lead to more gentle gas accretion onto the BH.


\begin{figure}
\begin{center}
\includegraphics[scale=0.4]{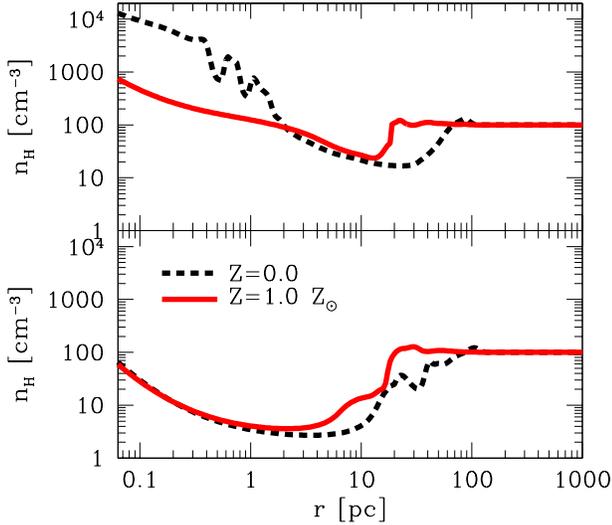}
\caption{ Density profiles in the models of M5-Z0 (black dash lines) and
  M5-Z1rad (red solid lines).  The upper panel shows snapshots near
  the peak of accretion rate, while the lower panel shows a
  snapshot when the accretion rate is low (quiescent phase).}
\label{fig:prof}
\end{center}
\end{figure}

Figure~\ref{fig:prof} presents the density profiles at the high and
low luminosity phases for runs M5-Z0 and M5-Z1rad.  At the high luminosity
phase, the density of M5-Z0 run steeply increases at $r \lesssim 1~\rm
pc$, and it reaches $\nh \sim 10^{4}~\rm cm^{-3}$ at the inner
boundary.  The inflow velocity exceeds sound speed at $\sim \rb / 2$,
and hence the density profiles are roughly $\propto r^{-3/2}$ as in
the Bondi profile (free falling gas).

On the other hand, the density of M5-Z1rad slowly increases as the
radial distance decreases.  In the case with dust, the net inward
force is significantly reduced due to the radiation force.  By
introducing an effective gravitational constant as 
$G' = G \left(1 - \fdust \fedd\right)$,
we estimate the transonic radius
by $r_{\rm t} \sim G' \Mbh / 2 \cs^{2}$.  When $\fedd \gtrsim
10^{-3}$, the transonic radius is smaller than the inner boundary of
the calculation box.   
  As a result, the gas inflow of M5-Z1rad is always
sub-sonic in our calculation box.  In addition, the temperature at $r
\lesssim 0.1~\rm pc$ increases as the radial distance decreases due to
compression heating.  Therefore, the pressure gradient force also
works on suppressing the gas inflow.  Thus the gas density does not
increase steeply.

As shown in the lower panel of the figure, the gas density is
decreased due to the photo-ionization feedback during the quiescent
low-luminosity phase.  At the ionization front, the gas is almost in
pressure equilibrium with the gas outside the H{\sc i} region, {
  i.e.}, $2 n_{\rm HII} T_{\rm HII} \sim n_{\rm HI}T_{\rm HI}$, where
the factor $\sim 2$ is due to the increased particle number from
ionization.  Since the temperature of the H{\sc ii} region reaches
$\sim 7 \times 10^{4}~\rm K$ via photo-ionization of hydrogen and helium
\citep{Park11}, the density of the H{\sc ii} region is roughly lower
than the ambient H{\sc i} gas by a factor $\sim 14$.  The gas density
is $\lesssim 100~\rm cm^{-3}$ even at the inner boundary in the
simulation.  Therefore the optical depth by dust inside the H{\sc ii}
region does not exceed unity as will be discussed in
Sec.~\ref{sec:rhii}.

\begin{figure}
\begin{center}
\includegraphics[scale=0.4]{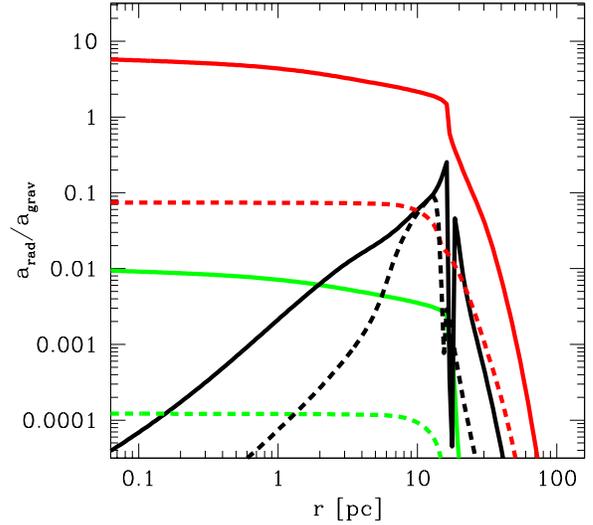}
\caption{ Radiation force normalized to the gravitational one as a
  function of radial distance for the M5-Z1rad run.  Black, green and red
  lines show the radiation force on H{\sc i}, electron and dust.  
  The solid lines refer to the burst phase
  ($\fedd=1.1\times10^{-2}$) and the dashed lines to the quiescent
  phase ($\fedd=1.2\times10^{-4}$).  }
\label{fig:prad}
\end{center}
\end{figure}
Figure~\ref{fig:prad} shows the gas outward acceleration due to
radiation pressure normalized by the gravitational acceleration.  Both
$\fradd$ and $\frade$ are roughly constant inside the H{\sc ii}
region.  The number ratio of dust grains to free electrons is almost
constant inside the H{\sc ii} region.  Therefore, the $\fradd$ is
higher than $\frade$ by a factor $\sim 700$ as shown by
Eq.~(\ref{eq:fratio}).  During the burst of accretion the dust
optical depth exceeds unity inside the H{\sc ii} region, because the
gas density increases near the BH as shown in Figure~\ref{fig:prof}.
Therefore the radiation pressure decreases as the distance increases
due to the dust opacity.  $\frade$ sharply drops outside the ionizing front
because the electron abundance decreases, while $\fradd$ gradually
decreases since dust exists both in the ionized and neutral gas.
 The radiation force on  H{\sc i}, $\fradhi$, monotonically increases with
 the radial distance in the H{\sc ii} region.  Where $\tau < 1$,
 $\fradhi / \fgrav$ is simply proportional to the neutral fraction
 $\xhi$.  Inside the H{\sc ii} region, hydrogen is in ionization
 equilibrium:
\begin{equation}
\int \frac{x_{\rm HI}L_{\rm \nu}
   \sigma_{\rm \nu}e^{-\tau_{\rm \nu}} }{4 \pi r^{2} h \nu} d\nu \sim
 \alpha_{\rm B} \nh (1-x_{\rm HI})^{2}.
\end{equation}

For $\tau \ll 1$ and $x_{\rm HI} \ll 1$, $x_{\rm HI} \propto \nh
r^{2}$.  $\fradhi$ increases somewhat more slowly than $r^{2}$. This is due
to the decrease of gas density as the radial distance increases at $r
\lesssim 10~\rm pc$.  Near the ionization front, $\fradhi$ steeply
increases with increasing $\xhi$, and decreases exponentially in the neutral region.
\footnote{
We notice that there is a dip in $\fradhi$ near the ionizing front. 
 At this radius, the flux steeply decreases as ${\rm exp}(-N_{\rm HI} \sigma_{\rm \nu, HI})$, 
while the neutral fraction increases at $\sim 1-2$ cells behind, producing the dip. 
This is due to the finite resolution. Increasing the number of cells makes this dip smaller. 
Since the radiation force on H{\sc i} is negligible, this feature does not affect our results.
}


\begin{figure}
\begin{center}
\includegraphics[scale=0.4]{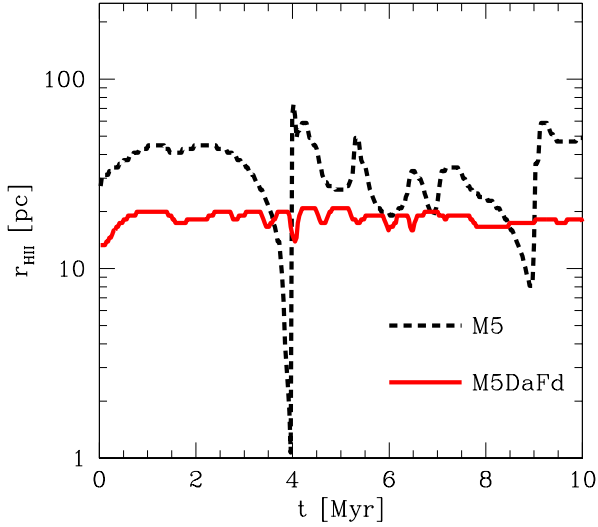}
\caption{ Time evolution of the size of ionized bubbles for two
  fiducial runs.  Black dash and red solid lines represent the bubble
  sizes in the M5-Z0 (without dust) and M5-Z1rad (with dust) runs, respectively.  }
\label{fig:rhii}
\end{center}
\end{figure}

\subsection{Size of ionized bubble}
\label{sec:rhii}
The time evolution of the position of the ionization front (defined
where $x_e=50~\%$) is presented in Fig.~\ref{fig:rhii}. The size of
ionized bubble $\rhii$ changes with the luminosity in the case of the
M5-Z0 run without dust.  When the pressure equilibrium at the ionization front breaks
due to the decreased density and luminosity, the neutral gas free falls into the
BH.  In this case, the inflowing gas remains neutral until
reaching near the BH, resulting in the collapse of the ionized bubble
just before the next bursts of luminosity.

\begin{figure}
\begin{center}
\includegraphics[scale=0.4]{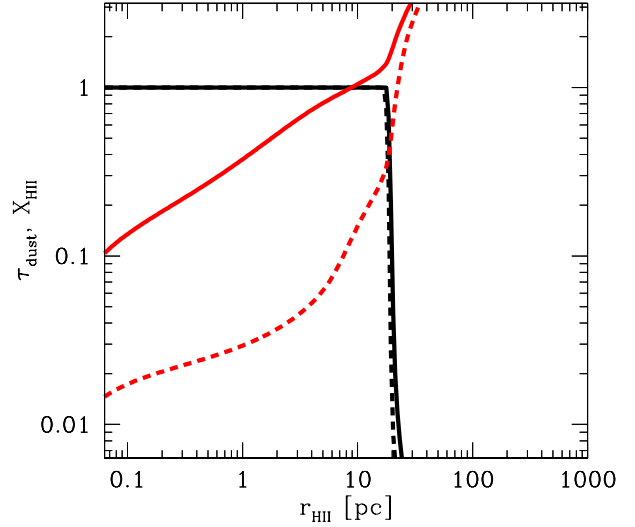}
\caption{ Optical depth of dust and hydrogen ionization fraction as a
  function of radial distance.  Red solid and dash lines show the
  optical depth during the high ($\fedd=1.1\times10^{-2}$) and low
  ($\fedd=1.2\times10^{-4}$) accretion phases, respectively.  Black
  solid and dash lines show the ionization fraction during the same
  high and low accretion phases.  }
\label{fig:taudust}
\end{center}
\end{figure}

On the other hand, in the dusty gas simulation (M5-Z1rad) the size of
the ionized bubble remains nearly constant, with $\rhii \lesssim
20~\rm pc$.  Even during the high luminosity phases, the size is
several times smaller than that of the dust-free case (M5-Z0).  This is
because dust opacity regulates the evolution of the size of the ionized bubble.
Figure~\ref{fig:taudust} shows the optical depth of dust during the high
and low luminosity phases. When the luminosity is high, the optical
depth of dusty gas reaches unity at $r \sim 10~\rm pc$ (M5-Z1rad).  Therefore the ionizing
front cannot propagate far beyond $10~\rm pc$, whereas in the dust-free case
(M5-Z0) the ionizing front (I-front) reaches $\sim 40~\rm pc$.  During the low luminosity phase, the
optical depth does not exceed unity in the ionized bubble due to the
lower gas density as shown in Fig.~\ref{fig:prof}.  In this case, the
size of ionized bubble is determined by the ionizing luminosity.  The gas
density steeply increases over the transition from ionized to neutral
regions.  This density jump significantly increases the optical depth
due to the high-density dust in the neutral region.  Therefore,
regardless of different luminosities, the optical depth exceeds unity
just behind the ionizing front.

\begin{figure}
\begin{center}
\includegraphics[scale=0.4]{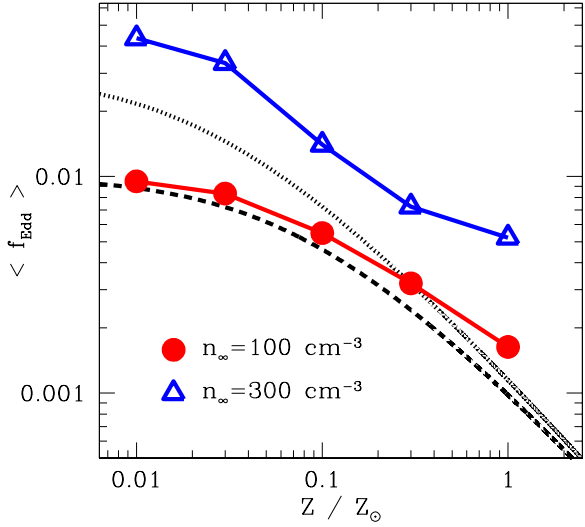}
\caption{ Time-averaged Eddington ratios as a function of metallicity.
  Red filled circles and blue open triangles show the Eddington ratios
  for initial gas densities of $100$ and $300~\rm cm^{-3}$,
  respectively.  Dashed and dotted lines represent the values
  estimated using Eq.~(\ref{eq:ana_metal}) for $\nhinf=100$ and
  $300~\rm cm^{-3}$, respectively.  }
\label{fig:mdot_D}
\end{center}
\end{figure}
 
\subsection{Metallicity dependence}

As star formation proceeds, the interstellar medium in galaxies becomes metal/dust enriched
by type-II supernovae \citep[e.g.,][]{Wise12a}.  Recent observations
indicated a large dispersion of dust amount in high-redshift galaxies
\citep{Watson15}.  Here we study the dependence of the BH accretion
luminosity on the gas metallicity and therefore the dust abundance
(since we assume a constant dust-to-metal mass ratio).
Figure~\ref{fig:mdot_D} shows $\feddt$ as a function of metallicity.
As the metallicity increases, $\feddt$ monotonically decreases, and
ranges from $\feddt \sim 10^{-2}$ at $Z=10^{-2}~\Zsun$ to $\feddt \sim
10^{-3}$ at $Z=\Zsun$.  We examine the metallicity dependence based on
the assumption of steady spherical inflow of ionized gas ({ i.e.,}
Bondi accretion inside the ionized bubble).  In the case without dust,
we estimate the accretion rate of Bondi-like inflow in the ionized
region ($\dot{M}_{\rm B, HII}$) as follows:
 \begin{equation}
  \dot{M}_{\rm B, HII} \sim 4 \pi \lambda_{\rm B} G^{2} \Mbh^{2} \rho_{\rm HII} c_{\rm s, HII}^{-3},
  \label{eq:mb}
 \end{equation}
 where $\lambda_{\rm B}$ is the dimensionless mass accretion rate,
 $\rho_{\rm HII}$ and $c_{\rm s, HII}$ are the density and the sound
 speed of ionized gas.  The value of $\lambda_{\rm B}$ depends on the
 polytropic index $\gamma$ of the equation of state, { i.e.},
 $P=K\rho^{\gamma}$, and ranges from $\sim 1.12$ for an isothermal gas
 to $1/4$ for the adiabatic case.  When we assume that the gas inflow
 is isothermal and consists of ionized hydrogen and helium with the
 temperature $T_{\rm HII} \sim 7 \times 10^{4}~\rm K$, the accretion
 rate from Eq.~(\ref{eq:mb}) is $\sim 770$ times smaller than the
 Bondi rate calculate from the density and temperature of the ambient
 medium. This reduction of the accretion rate is roughly similar to our
 simulation results and previous works \citep{Milosavljevic09a,
   Park12, Sugimura16}.  Note that, however, the inflow of primordial
 gas causes periodic bursts as shown in the Sec.~\ref{sec:mdot}.
 These bursts can increase the time averaged accretion rate with respect to
 that estimated above \citep[see also,][]{Park12}.
 
 Next, we modify the above accretion rate by taking into account the
 radiation pressure on dust.  Due to radiation pressure, the net
 inward force is reduced.  In the optically thin regime ($\tau_{\rm d}
 < 1$), the fractional reduction of the inward gravitational force is
 independent to the radial distance, and depends only on the
 BH luminosity.  In this case, the accretion rate is
 \begin{equation}
 \begin{split}
 \dot{M} \sim 4 \pi \lambda_{\rm B} G^{2} \left( 1 -  \frac{f_{\rm d} \sigma_{\rm T} L}{4 \pi c G \Mbh m_{\rm p}} \right)^{2}
  \Mbh^{2} \rho_{\rm HII} c_{\rm s, HII}^{-3}, \\
  = \dot{M}_{\rm B, HII} \left[ 1 -  \frac{7.1\times10^{2} \sigma_{\rm T} \eta c }{4 \pi G \Mbh m_{\rm p}} \left( \frac{Z}{\Zsun} \right) \dot{M} \right]^{2}.
 \end{split}
 \label{eq:ana_metal}
 \end{equation}
By solving this second order equation, we derive an analytical
estimate of the accretion rate and luminosity.  The analytically
estimated Eddington ratios are shown as dashed ($\nhinf=100~\rm
cm^{-3}$) and dotted ($\nhinf=300~\rm cm^{-3}$) lines in the
Figure~\ref{fig:mdot_D}.  As shown in the figure, the modified Bondi
rate considering the radiation force roughly explains the metallicity
dependence of the accretion rate in the simulations with radiation
pressure on dust.  Note that, $\fedd$ in the high metallicity and high
density cases differ from the value estimated above, whereas it in the
case with $\nhinf = 100~\rm cm^{-3}$ our analytical estimate is good
fit to the simulation results.  When the density and metallicity are
high, the size of H{\sc ii} bubble is close to the Bondi radius \citep[e.g.,][]{Park12}.
Therefore, the gas density in the ionized region can become somewhat
higher than that estimated assuming pressure equilibrium across the
ionization front due to the BH gravitational potential, resulting in
higher accretion rate.
 
\subsection{Mass dependence}

\begin{figure}
\begin{center}
\includegraphics[scale=0.4]{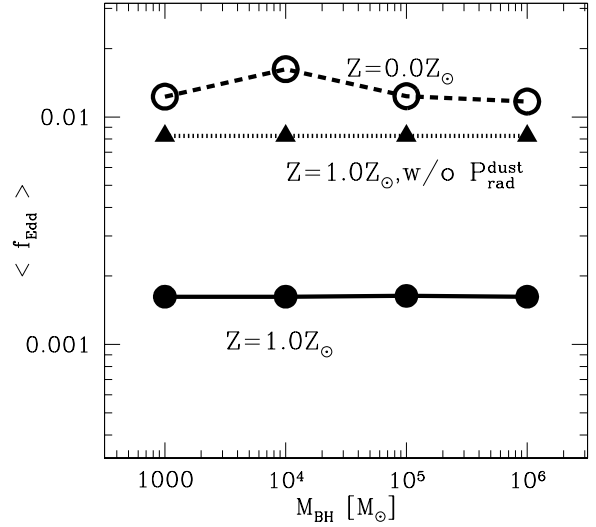}
\caption{ Time-averaged Eddington ratio as a function of BH mass.
  Filled and open circles show the cases with dust ($Z=1.0~\Zsun$),
  and without dust ($Z=0.0~\Zsun$), respectively.  Filled Triangles
  represent the case with dust opacity, but without radiation force on
  dust.  In these runs the gas density changes with the BH mass under the
  constraint $\nh = 100 \left( 10^{5}~\Msun / \Mbh \right)~\rm
  cm^{-3}$.  }
\label{fig:mdot_mass}
\end{center}
\end{figure}

Here we study the BH mass dependence on the gas accretion rate for our
fiducial runs with $\Mbh \nh=10^{7}~\rm \Msun ~cm^{-3}$.
\citet{Park12} showed that the accretion rate normalized by the Bondi
rate did not change for different BH masses when $\Mbh \nh=10^{7}~\rm
\Msun ~cm^{-3}$.  Figure~\ref{fig:mdot_mass} shows the time averaged
Eddington ratios for different BH masses.  As shown in \citet{Park12},
$\feddt \sim 1\%$ independently of the BH mass in the case without
dust.  We find that the same result holds even when including
dust. Thus, regardless of the dust and metallicity of the gas, we
confirm that the results are scale-free under the condition $\Mbh \nh
= 10^{7}~\rm \Msun ~cm^{-3}$.

\subsection{Thermal emission from dust}

\begin{figure}
\begin{center}
\includegraphics[scale=0.4]{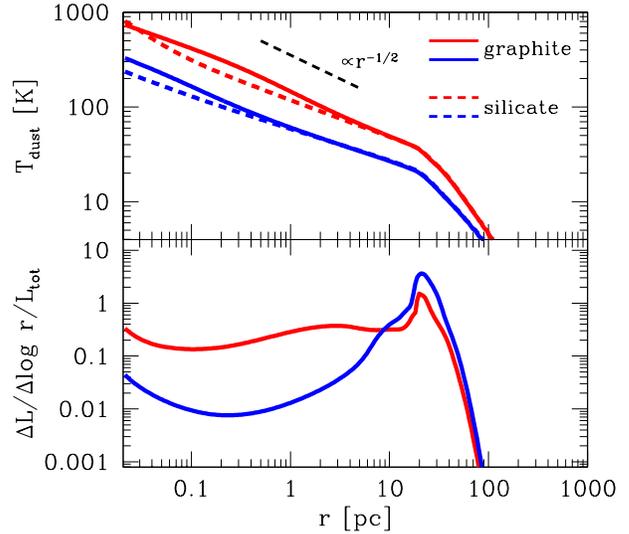}
\caption{ Upper panel: Dust temperature profile for the M5-Z1rad run.
  Solid and dash lines represent the temperatures of graphite and
  silicate dust grains, respectively.  The red and blue colors show
  the high ($\fedd=1.1\times10^{-2}$) and low
  ($\fedd=1.2\times10^{-4}$) luminosity phases.  Lower panel: The
  energy fraction absorbed by each spherical shell for the same run
  and accretion phases as in the top panel.  }
\label{fig:tdust}
\end{center}
\end{figure}

Dust releases the energy absorbed at UV wavelengths as thermal
emission in the IR. Here we investigate the IR properties
of the dusty gas accreting onto the BH.  The dust temperature $T_{\rm d}$ can be
estimated by assuming radiative equilibrium:
\begin{equation}
\int \pi \ad^{2} Q_{\rm \nu} F_{\rm \nu} \; d\nu = \int 4 \pi \ad^{2} Q_{\rm \nu} \pi B_{\rm \nu} (T_{\rm d}) d\nu.\label{eq:td}
\end{equation}
This equation assumes equilibrium between dust absorption of the radiation emitted by the BH (left-hand side) and the thermal emission from dust (right-hand side).
By solving this equation, we estimate the dust temperature of each gas
shell.  We assume $Q_{\nu} = 1$ at UV wavelengths for the left-hand
side of the equation, and use $Q_{\nu}$ estimated in \citet{Laor93}
for the right-hand side.
 In order to estimate the thermal
  emission by dust we post-process the simulation results, calculating
  the absorbed flux by dust.  In the estimation of the dust thermal
  emission, we neglect the absorption of ionizing photons by hydrogen and
  helium.  This is because the absorbed energy is converted into
  recombination and cooling radiation, e.g., Ly$\alpha$ photons.
  These recombination UV photons are eventually absorbed by dust.  In
  addition, by considering only dust absorption, the IR properties do
  not change significantly even if we consider different choices for
  the SED from the accretion disk, e.g., a SED extending to soft-UV
  range.   Figure~\ref{fig:tdust} shows the dust temperature as a
function of distance from the BH.  During the high-accretion phase,
the dust temperature is $\sim 800~\rm K$ near the inner boundary of
the simulation, and decreases as the distance increases due to the
geometrical attenuation of the flux $\propto r^{-2}$.  Due to the
strong attenuation of UV flux outside the I-front, the dust
temperature sharply drops at $r \gtrsim 10~\rm pc$.  In regions where
$\td \gtrsim 100~\rm K$, the dust temperature of silicate is somewhat
lower than graphite.  This is because $Q_{\rm \nu}$ of silicates is
higher than graphite in that temperature range, thus the efficient
thermal photon emissivity results in the lower temperature.

In order to better understand the radial profile of the dust temperature, we
re-write Equation~(\ref{eq:td}) as:
\begin{equation}
F \sim 4 \sigma_{\rm SB} T_{\rm d}^{4} \bar{Q}(T_{\rm d}),
\end{equation}
where $F = \int F_{\nu} d\nu$ is radiation flux and
$\bar{Q}(T_{\rm d}) = {\int B_{\nu}(T_{\rm d}) Q_{\rm \nu} d\nu}/
{\int B_{\nu}(T_{\rm d}) d\nu}$ is the frequency-averaged dust
absorption coefficient.
Therefore, the dust temperature is
\begin{equation}
T_{\rm d} (r) =  \left[
\frac{L \; {e}^{-\tau_{\rm d}(r)}}
{16 \pi \sigma_{\rm SB} \bar{Q}(T_{\rm d}) r^{2}}
\right]^{\frac{1}{4}}. 
\end{equation}
For $\tau \ll 1$, this equation is scaled as follows:
\begin{equation}
\begin{split}
T_{\rm d}(r) = 4.7 \times 10^{2}~{\rm K}~
\left( \frac{\fedd}{10^{-2}} \right)^{\frac{1}{4}} 
\left( \frac{\Mbh}{10^{5}~\Msun} \right)^{\frac{1}{4}} \\
\times \left( \frac{\bar{Q}}{10^{-2}} \right)^{-\frac{1}{4}}
\left( \frac{r}{0.1~\rm pc} \right)^{-\frac{1}{2}}. 
\end{split}
\label{eq:tdust}
\end{equation}
As shown by the figure, the dust temperature roughly decreases as
$\propto r^{-1/2}$.  However, note that, $\bar{Q}$ decreases as the
dust temperature decreases.  Therefore, the radial profile of the dust
temperature is somewhat shallower than $r^{-1/2}$.

The lower panel of Fig.~\ref{fig:tdust} shows the radial dependence of
the fraction of energy absorbed by the dust per unit radial log bin,
normalized by the total luminosity.  During the quiescent accretion
phase, most of energy is absorbed by dust near the ionizing front.
This is because the optical depth at $r < 10~\rm pc$ is much smaller
than unity as shown in Fig.~\ref{fig:taudust}.  Therefore, most of IR
emission comes from cold dust with $\tdust \sim 20~\rm K$ near the
ionizing front.  As the accretion rate increases, a fraction of the
radiation is absorbed by dust near the inner boundary of calculation
box due to higher gas and dust density.  This results in a large
contribution to the IR luminosity by the inner hot dust with $\tdust
\gtrsim 100~\rm K$.

\begin{figure}
\begin{center}
\includegraphics[scale=0.4]{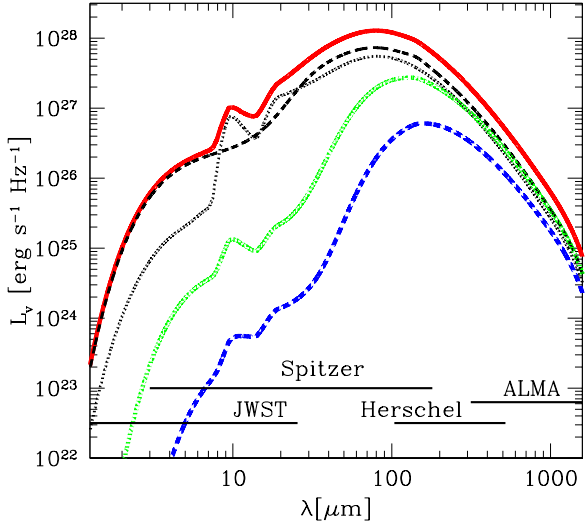}
\caption{
SED of hot and cold dust at IR wavelengths around the
  $10^5$~M$_\odot$ BH in the M5-Z1rad run.  Red solid, green dot, and blue
dash lines show the specific luminosities during high
($\fedd=1.1\times10^{-2}$), moderate ($\fedd=0.8\times10^{-3}$), and
low ($\fedd=1.2\times10^{-4}$) luminosity phases.  Black dash and dot
lines show the contributions from graphite and silicate dust grains to
the specific luminosity during the high luminosity phase (red
solid line).  }
\label{fig:sed}
\end{center}
\end{figure}

By integrating the thermal emission of each spherical shell, we derive
SEDs. 
Figure~\ref{fig:sed} shows the SEDs at $\fedd=1.1 \times10^{-2}$ (red
line), $0.8 \times10^{-3}$ (green line), and $1.2 \times10^{-4}$ (blue
line).  The red and blue lines correspond to the ones in
Fig.~\ref{fig:tdust}.  We also show the frequency ranges of Spitzer, JWST,
Hershel and ALMA.  Due to the contribution to the IR luminosity by hot
dust, the SED during the high accretion phase shows stronger IR
emission at $\lambda \lesssim 50~\rm \mu m$.  The specific luminosity
at these wavelengths varies by several orders of magnitude between the
quiescent phase with low Eddington ratio and the bursty phase at
higher Eddington ratio. On the other hand, at $\lambda \gtrsim 100~\rm
\mu m$ the specific luminosity remains nearly constant during the
accretion cycle, because the cold dust with $T_{\rm d} \lesssim 50~\rm
K$ near the ionization front, always absorbs and reprocesses a large
fraction of the radiation.
 Recently, \citet{Shimizu17} showed
  that the IR SEDs of AGNs selected from their hard X-ray luminosity
  using Swift BAT, correlates with the X-ray luminosity $L_{\rm X}$:
  with increasing $L_{\rm X}$, the flux at short wavelength $\lambda
  \sim 10~\rm \mu m$ also increases. If the BH masses of the observed
  AGNs in the sample are similar each other, these variations of the
  SEDs can be explained by the different phases of accretion and
  the Eddington ratios as explained above. An estimate of the masses and Eddington ratios of
  these hard X-ray selected AGNs is currently in progress (T. Shimizu,
  private communication). With the caveat of being able to isolate the AGN contribution from the stellar contribution in the IR SEDs, the results of this study can be used to test our model predictions.  

We also show the contributions of graphite and silicate grains,
separately.  The stretching resonance of astronomical silicate grains
produces the bump in the SED at $\lambda \sim 9.8~\rm \mu m$.  At
shorter wavelengths ($\lambda \lesssim 8 ~\rm \mu m$), the
contribution from graphite grains is dominant.  In this work, we have
not considered re-absorption of the thermal emission by dust, because
the absorption cross-section of dust to IR photons is quite small and
we assume that the ambient gas is optically thin at IR wavelengths.
However, if the hydrogen column density of the gas reservoir fueling
the BH is much higher than $10^{22}~\rm cm^{-2}$, some IR radiation
emitted by hot dust can be absorbed by the dust and re-processed to
thermal emission at lower dust temperature.  This can somewhat
suppress the IR flux at short wavelength.

\begin{figure*}
\begin{center}
\includegraphics[scale=0.9]{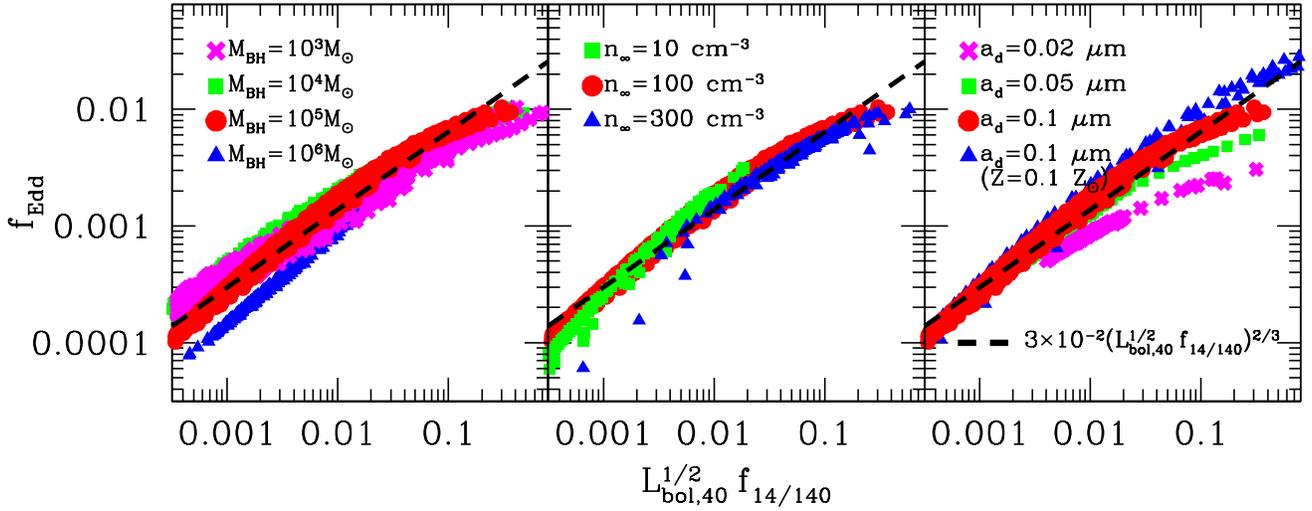}
\caption{
Eddington ratio as a function of $L_{\rm bol, 40}^{1/2} f_{14/140}$, 
where $L_{\rm bol, 40} (\equiv L_{\rm bol}/10^{40}~\rm erg~s^{-1})$ is an IR bolometric luminosity normalized by $10^{40}~\rm erg~s^{-1}$
and $f_{14/140}$ is the flux ratio between $14$ and $140~\rm \mu m$.
Different symbols in the left panel refer to simulations with different BH masses but  $\Mbh \nh ={\rm
    const.}$
The middle panel shows the Eddington ratios for different initial gas densities.
The right panel shows simulation with different dust size models. 
Only the blue triangles refer to simulations with metallicity $Z=0.1~\Zsun$ and fiducial dust size $0.1~\rm \mu m$. All other simulations assume solar metallicity. The black dashed line is the power-law function roughly reproducing our simulations results.
}
\label{fig:Lfratioall}
\end{center}
\end{figure*}

  Fig.~\ref{fig:Lfratioall} shows the relation between the
  Eddington ratio, $\fedd$, and the flux density ratio between $14$
  and $140~\rm \mu m$ ($f_{14/140}$) multiplied by the square root of
  bolometric luminosity, $L_{\rm bol, 40}$ (in units of
  $10^{40}~\rm erg~s^{-1}$).  As shown in Fig.~\ref{fig:sed}, the flux
  at $\lambda \lesssim 10~\rm \mu m$ depends sensitively on the
  presence of a hot dust component.  Therefore we choose the flux at
  $14~\rm \mu m$ as diagnostic of the high-accretion phase.  The
  wavelengths $14$ and $140~\rm \mu m$ correspond to the peak emission
  of two black body spectra with temperatures of $207.0$~K and
  $20.7~\rm K$, respectively.  We find that this flux ratio tightly
  correlates with $\fedd$.
The correlation between $\fedd$ and $f_{14/140}$ varies with the BH
mass, and dust size, mainly because the
temperature of the warm and hot dust components depend on these
physical parameters.  However, $\fedd$ tightly correlates with $f_{14/140}$ regardless of the assumed initial background density.
Yet, we can use our knowledge of $L_{\rm bol, 40}$ to correct
for these effects and recover $\fedd$ regardless of the different
physical parameters of the SMBH, that are generally unknown
observationally.
The different panels in the figure illustrate the tight correlation
between $\fedd$ and $L_{\rm bol, 40}^{1/2}f_{14/140}$, that holds by varying
different physical parameters by many orders of magnitude. We find that the
correlation can be roughly fitted by the following relationship:
\begin{equation}
  \fedd \sim 3 \times 10^{-2} \left( L_{\rm bol, 40}^{1/2} f_{14/140} \right)^{2/3}.
  \label{eq:fedd}
\end{equation}
Therefore, keeping into account that our model is still very simplistic, it may be possible to use IR observations of warm and hot dust obscuring SMBHs to estimate $L_{\rm bol, 40}$ and $f_{14/140}$, and therefore derive $\fedd$ and the SMBH mass as
\begin{equation}
\Mbh \sim 8 \times 10^{3}~\Msun~L_{\rm bol, 40} \left(\frac{\fedd}{10^{-2}}\right).
\end{equation}

Since $L_{\rm bol, 40} \equiv \fedd L_{\rm Edd} \propto \fedd
  \Mbh$, we can use Eq.~(\ref{eq:fedd}) to show that $\fedd \propto
  \Mbh^{1/2} f_{14/140}$. Therefore, for a fixed BH mass, we indeed
  find $\fedd \propto f_{14/140}$.  Instead, keeping constant $\Mbh
  \nhinf$, while varying the density and/or BH mass, we find $\rhii
  \propto \fedd^{1/3} \Mbh$ (see Eq. \ref{eq:rhii}).  Using
  Eq.~\ref{eq:tdust}, the dust temperature at $\rhii$ is $\td(\rhii)
  \propto \Mbh^{-1/4}$ (here we ignore a weak dependence on $\fedd$).
  Therefore, the weak mass dependence in the figure indicates
  $f_{14/140}$ decreases as $\td(\rhii)$ decrease with increasing the BH mass (and $L_{\rm bol, 40}$).  

We also study the SEDs for different dust size models, $a_{\rm d}=0.05$ and $0.02~\rm \mu m$.  
The right panel of Fig.~\ref{fig:Lfratioall} shows the correlations in the cases of different dust sizes. 
Unlike the tight correlations found for different BH masses and densities, 
models with the different dust sizes produce a systematic change in shape of the relationship.
Since $Q_{\rm abs}$ at IR band decreases with dust size, smaller dust grains become hotter when irradiated by the same UV flux as shown in Eq.~(\ref{eq:tdust}). Moreover, the size of the H{\sc ii} region becomes smaller as the dust size decreases (see Eq.~(\ref{eq:taudust})). Thus, models with smaller dust tend to have higher $f_{14/140}$ for a fixed $\fedd$. In the same panel we also 
show the correlation for the low-metallicity (low-dust) case with $Z=0.1~\Zsun$. 
As the metallicity decreases, the BH luminosity and the ionize bubble radius, $\rhii$, increase.
As a result, the dust temperature does not change significantly, producing the same correlation as in the fiducial case with $Z=1~\Zsun$.   
Therefore we suggest the uncertainty of metallicity is not needed to be considered. 

Can JWST or ALMA observe thermal emission from the dust obscuring massive BHs in nearby galaxies?
For low-redshift AGNs, the flux from hot dust at rest-frame $\lambda
\sim 14~\rm \mu m$ can be observed by JWST, while the warm dust
emission at $\lambda \sim 140~\rm \mu m$ is in the wavelength range
covered by Spitzer.
Even if we measure the flux density ratio between $14$ and $450~\rm \mu m$ (instead of $140~\rm \mu m$) that is in the waveband covered by ALMA, 
as similar relationship between $\fedd$ and the flux ratio is found. 
Therefore ALMA can also be used to trace the cold dust at $z \sim 0$.

\begin{figure}
\begin{center}
\includegraphics[scale=0.53]{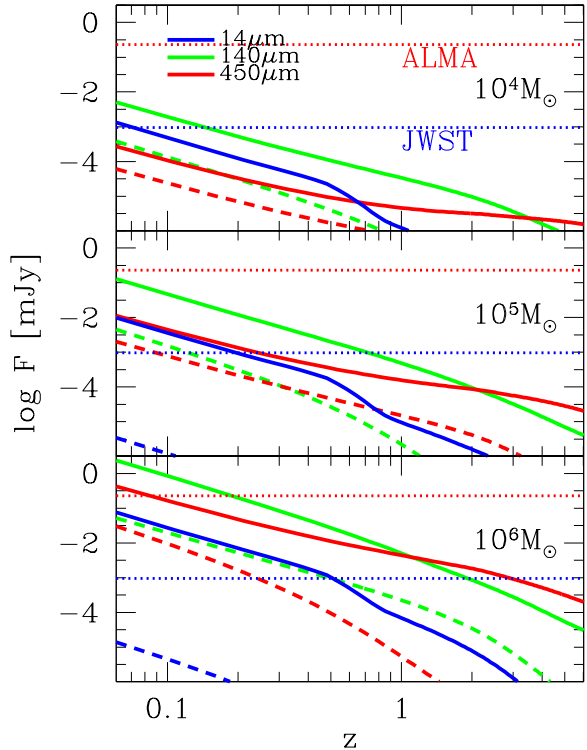}
\caption{
Observed flux from thermal emission of hot and warm dust around accreting BHs as a function of redshift.
Different colors lines show different wavelengths in the observer rest frame: $14~\rm \mu m$ (red), 
$140~\rm \mu m$ (green), $450~\rm \mu m$ (red).
Solid and dashed lines refer to the peak accretion rate ($\fedd \sim 10^{-2}$) and the quiescent phase ($\fedd \sim 10^{-4}$), respectively.
Horizontal dotted lines show $10~\sigma$ detection limits with 10 hours integration at $14~\rm \mu m$ by JWST (F1500W filter) and at $450~\rm \mu m$ by ALMA with 50 antennas. 
}
\label{fig:fobs}
\end{center}
\end{figure}

 Using the specific luminosities derived above, we estimate the
  flux densities from dust around massive BHs at a specific redshift
  $z$ as $F_{\rm \nu} = (1+z)L_{\rm \nu 0} / 4 \pi D_{\rm L}^{2}$
  where $D_{\rm L}$ is the luminosity distance, $\nu$ is the frequency
  in the observer rest frame, and $\nu_{0}=\nu(1+z)$ is the frequency
  in the galaxy's rest frame.  Here we artificially put BHs at $z =
  0.05 - 10$, and discuss their observability by JWST or ALMA.
  Figure~\ref{fig:fobs} shows the flux densities as a function of
  redshift. The dotted lines show the sensitivities of JWST and ALMA.
  We show the flux densities at $14~\rm \mu m$ ($F_{14}$), $140~\rm
  \mu m$ ($F_{140}$) and $450~\rm \mu m$ ($F_{450}$) in the observer
  frame.  Due to the negative K-correction, the flux density at
  $450~\rm \mu m$ decreases slowly with increasing redshift, whereas
  that at $14~\rm \mu m$ decreases more steeply at $z \gtrsim 1$.  At
  higher redshift, the flux at $14~\rm \mu m$ in observer rest frame
  corresponds to a wavelength in the galaxy's rest frame of $14~{\rm
    \mu m}/(1+z)$ which can be shorter than the peak wavelength of
  modified black body of even hot dust, resulting in the decreases of
  the flux.  As a result, the flux density at $450~\rm \mu m$ becomes
  higher than that at $14~\rm \mu m$ at $z > 1$.  The relation between
  $F_{14}$ and $F_{450}$ changes depending on the BH mass and $\fedd$.
  As shown above, $F_{14}$ is lower than $F_{140}$ or $F_{450}$ when
  $\fedd$ is low.  Even at high-accretion phase ($\fedd \sim
  10^{-2}$), $F_{450}$ is higher than $F_{14}$ in the case of massive
  BH with $10^{6}~\rm \Msun$.  Under the constraint $\Mbh \nh ={\rm
    const}$, we have $\rhii \propto \Mbh$ (see Eq. \ref{eq:rhii}).
  Therefore, the dust temperature at the ionization front is lower in
  massive BHs (see Eq. \ref{eq:tdust}), leading to a higher $F_{450}$.

In summary, we find that only massive BH of mass $\gtrsim 10^{6}~\rm \Msun$ at $z \lesssim 0.1$ can be observed by ALMA.
On the other hand, JWST will allow us to observe the dust thermal emission from more distant massive BHs. 
It will be able to probe hot dust obscuring BHs of $10^{6}~\Msun$ with $\fedd \sim 10^{-2}$ up to $z \sim 0.5$. 
When $\fedd \sim 10^{-2}$, even a $10^{4}~\Msun$ BH at $z \lesssim 0.1$ can be observed. 
Thus we suggest JWST and ALMA will be powerful tools to probe IMBHs via the observation of dust thermal emission.
In general, the IR flux increases as the product $\Mbh \nhinf$ increases. For instance, for the case of a $\Mbh=10^{6}~\Msun$ accreting from gas with density $\nhinf=100~\rm cm^{-3}$ and metallicity $Z=0.01~\Zsun$, ALMA and JWST can probe BHs up to $z \sim 0.5$. However, even upcoming new telescopes will be difficult to observe IMBHs at $z \gg 1$ in the IR.

In our simulations, we have investigated the accretion dynamics of spherically symmetric clouds
with isotropic radiation field
in which the accretion rate and luminosity can be suppressed significantly.
Recently, \citet{Sugimura16} suggested the gas efficiently accreted onto a BH if the radiation
is anisotropic due to shadowing effects near the edge of the BH accretion disk or torus.
In this case, $\fedd$ can become $\gtrsim 1$, and even relatively massive BHs at high-redshifts $z \gtrsim 1$ may be observable by JWST or ALMA, although the dust temperature and SEDs can differ from our current works.


%
%

\section{Discussion} 
\label{sec:discussion}

\subsection{Dust decoupling}
In this work we have assumed that the motion of dust is completely
coupled to the gas through collisions between gas and dust particles.
Here we simply estimate the coupling time scale.  At first, we
consider the momentum equation for the relative velocity $v$ between dust and
gas:
\begin{equation}
\frac{dv}{dt} = - n_{\rm H}\frac{m_{\rm H}}{m_{\rm d}} \pi a_{\rm d}^{2} v^{2}.
\end{equation}
Therefore, the time scale for the dust coupling with gas is roughly
\begin{equation}
\begin{split}
\tau &\sim \frac{v}{|dv/dt|} = \frac{4}{3} \frac{\rho_{\rm d} a_{\rm d}}{n_{\rm H}m_{\rm H}v} \\
&= 6.0\times10^{3} \; {\rm yr} \left( \frac{\ad}{0.1\;{\rm \mu m}}\right)
\left(  \frac{\nh}{10^{2}\; \rm cm^{-3}} \right)^{-1}
\left( \frac{v}{10\; \rm km\;s^{-1}} \right)^{-1}
\end{split}
\end{equation}
This coupling timescale is much shorter than dynamical timescale and
the period between bursts in our current simulations.  Thus the
assumption of the complete coupling is reasonable.

\subsection{Dust destruction}


The dust formation/destruction processes have not been considered in
our simulations.  Even at innermost cell ($=0.02~\rm pc$), the dust
temperature does not exceed the dust sublimation temperature.  For the
sublimation temperatures of graphite grains, $T_{\rm sub} \sim
1800~\rm K$, we estimate the destruction radius as follows:
\begin{equation}
\begin{split}
r_{\rm sub} = 6.8 \times 10^{-3} \; {\rm pc} \; \left( \frac{f_{\rm Edd}}{10^{-2}} \right)^{\frac{1}{2}} \\
\times \left(  \frac{\Mbh}{10^{5}~\Msun} \right)^{1/2} \left[ \frac{\bar{Q}(T_{\rm sub})}{0.2} \right]^{-1/2}.
\end{split}
\end{equation}
Thus, the dust can survive sublimation by photoheating even inside the
inner boundary of the calculation box.

On the other hand, the destruction by thermal sputtering process, {
  i.e.}, collisions between dust and gas, is likely to be effective
near the BH because of the high-density and temperature found at the
innermost cells in our simulations.  The destruction time scale by the
thermal sputtering is estimated \citep{Draine79, Draine11} as follows:
%
\begin{equation}
\tau_{\rm sp} \sim 1 \times 10^{5} \; {\rm yr} \; \left[1 + \left(\frac{T}{10^{6}~\rm K}\right)^{-3}\right]  \frac{(a_{\rm d}/ 0.1\rm \mu m)}{(n_{\rm H} / 1~{\rm cm^{3}})}.
\end{equation}
During the quiescent phase, the density, temperature and inflow
velocity at the innermost cell are $\sim 10^{3}~{\rm cm^{-3}},~ 3
\times 10^{5}~\rm K$ and $\lesssim 10 ~\rm km~s^{-1}$, respectively.
Therefore, during this phase in the duty cycle, the sputtering time
scale is shorter than the inflow time scale, $t_{\rm dyn} \sim
r/v_{\rm in}$ and the dust is likely to be destroyed inside the inner
boundary of the calculation box.  However, during the peaks of
accretion, the temperature somewhat decreases and the velocity
increases, resulting in the longer sputtering time scale. Therefore,
during the accretion burst dust can survive even close to a BH, and
can absorb UV radiation emitted during the burst. However, as shown in
Figure~\ref{fig:tdust}, the contribution to the opacity by dust
absorption at radii smaller than the inner boundary in our simulations
is not significant.  In addition, Compton heating increases the
temperature of gas at such the scale \citep{Park14b}, leading to
enhancement of dust destruction.  Yet, photo-ionization heating
of metals is not included in this work.  This heating can expand the
destruction radius.  These detailed heating processes will be
considered in our future works.

In this work, we adopt a single-size dust model in the calculations.
When considering a realistic dust size distribution, the destruction
radius depends on the dust size.  This is because small dust grains
can have higher temperature, hence they reach the sublimation
temperature at larger distance.  In addition, the destruction time
scale by thermal sputtering is proportional to the size.  Thus,
smaller dust grains are more easily destroyed near the BH compared to
larger dust grains.  This indicates that the dust size distribution
changes with radial distance.

\subsection{Revisited condition for hyper-Eddington accretion}

In this work we have investigated the density dependence up to
$\nh=1000~\rm cm^{-3}$.  This condition allows the ionizing front to
propagate far from the Bondi radius in the case without dust.
Recently, \citet{Inayoshi16} suggested that BHs could grow at
super-Eddington rate if they are embedded in very high-density gas
clouds where the size of ionized bubble is smaller than the Bondi
radius.  This is because the ionized bubble is gradually shrunk as the
gas density near the ionization front increases due to the
gravitational force by the BH.  Finally the ionized bubbles disappear,
resulting in no radiation force on electrons.  As a result the growth
rate of BH becomes close to the Bondi rate which can be much higher
than the Eddington limit, { i.e.}, so-called the hyper-Eddington
accretion.  They estimated the critical gas density for the
hyper-Eddington accretion by comparing the Str\"{o}mgren radius to the
Bondi radius. 
By calculating the Str\"{o}mgren radius for the density inside the ionized region (which is
smaller than $\nhinf$ by a factor $\sim 2 (T_{\rm HII}/ T_{\rm HI})$), we derive the critical density:
\[
\nhinf  \gtrsim 10^{5}~{\rm cm^{-3}} 
\left(
\frac{\Mbh}{10^{5}~\Msun} \right)^{-1} \left( \frac{T}{10^{4}~\rm K}
\right)^{3/2},
\]
where we have assumed $T_{\rm HII}=7 \times 10^{4}~\rm K$.
On the other hand, in the case of dusty gas, the size of ionized
region is regulated by the dust opacity if the metallicity is higher
than a critical value ({ i.e.}, if $Z > 0.1 Z_\odot$).  In this regime, the size of ionization region
is close to the photon mean free path, i.e., $\rhii \sim
(f_{\rm d} \sigma_{\rm T} n_{\rm HII})^{-1}$.  Thus, the critical
density for super-Eddington accretion in a dusty gas is:
\begin{equation}
\nhinf \gtrsim  2 \times10^{3}~{\rm cm^{-3}} \left( \frac{\Mbh}{10^{5}~\Msun}\right)^{-1} \left( \frac{T}{10^{4}~\rm K} \right) \left( \frac{Z}{Z_{\rm \odot}} \right)^{-1}.
\end{equation}
This critical density is much lower than that for the primordial gas
case. 
However, the effect of dust in
the regime of super-Eddington accretion, in addition to reducing the
size of the ionized bubble, is to reduce the accretion rate due to the
radiation pressure on dust. Radiation pressure is effective at radii larger than
the size of the ionization region, as long as $\tau_{\rm d} \lesssim
1$.  Therefore, if the ram pressure force can overcome the radiation
force on dust at the radius of $\tau_{\rm d} \sim 1$, the dusty gas may be able
to efficiently accrete onto the BH.  The accretion dynamics of
high-density dusty clouds will be investigated in future works with
further improvements to the code.

%
%

\section{Summary}
\label{sec:summary}

In this paper we have studied the accretion of dusty gas onto an
intermediate mass black hole (IMBH) by using one-dimensional radiation
hydrodynamics simulations.  \citet{Park11} showed that the growth of
BH was significantly regulated due to the photo-ionization feedback in
the case of primordial gas.  As star formation proceeds, high-redshift
galaxies are metal/dust enriched due to type-II supernovae.  Recent
observations have detected dust-rich galaxies even at $z \gtrsim 6$
\citep[e.g.,][]{Riechers13}.  This indicates some BHs grew in dusty
gas with the feedback.  In this paper, we investigate the effects of
dust on the growth of BH and observable diagnostics due to the thermal
emission from heated dust. Dust affects the accretion rate and period
of the bursts of accretion mainly because of radiation pressure on
dust but also because of the dust opacity that reduces the size of the
ionization region.

By assuming as a fiducial model accretion onto a BHs of $10^{5}~\Msun$
embedded in a uniform density medium, we investigate the dependence of
the BH growth rate on the gas density (in the range $\nhinf=10$ to
$1000~\rm cm^{-3}$), and on the metallicity (in the range $Z=0$ to $1~
\Zsun$).  We find that the accretion of dusty gas onto IMBHs proceeds
gently with small fluctuations of the accretion rate, whereas that of
primordial gas causes periodic bursts.  For dust-to-gas mass ratio
similar to the solar neighborhood, the time averaged luminosity
becomes smaller than that for primordial gas by one order of
magnitude.  The time averaged Eddington ratio is $\feddt \sim 10^{-3}$
for the initial gas density $\nhinf = 100~\rm cm^{-3}$.  Our
calculations show that the effect of dust opacity is secondary with
respect to radiation pressure on dust. Neglecting radiation pressure
on dust but including the effect of dust opacity, the growth rate of
IMBHs and $\feddt$ are smaller than that of primordial gas by a factor
$\lesssim 2$.  For both primordial and dusty clouds, the Eddington
ratio, $\feddt$, linearly increases with the initial gas density.  In
addition, assuming the constraint $(\Mbh/\Msun) (\nhinf/{\rm cm^{-3}})
= 10^{7}$, we study the dependence of the growth rate on the BH mass.
We show that $\feddt$ is constant for the different BH masses in the
both cases with and without dust.
 
Finally, we derive the SEDs at IR bands by calculating dust thermal
emission.  Our modeled SEDs show that the flux ratio at
$\lambda \lesssim 20~\rm \mu m$ and $\gtrsim 100~\rm \mu m$ depends
sensitively on the Eddington ratio, but is nearly independent of the
other parameters in the problem.  This is because at high Eddington
ratios the thermal emission from hot dust near the BH produces a higher
flux density at $\lesssim 20~\rm \mu m$. While the emission at
$\gtrsim 100~\rm \mu m$, that is produced by warmer dust further out, near the
ionization front, is nearly independent of $\feddt$.  Therefore, we
suggest that the combinations of MIR observations by JWST and FIR
observation by Hershel or ALMA can be combined to provide a novel
method to estimate the Eddington ratio of BHs throughout their duty
cycle, including their short bursty phase and their longer quiescent
phase.

%
%
\section*{Acknowledgments}
The numerical simulations were performed on the computer
cluster, {\tt Draco}, at Frontier Research Institute for
Interdisciplinary Sciences of Tohoku University.  This work is supported
in part by MEXT/JSPS KAKENHI Grant Number 15H06022 (HY) and 15J03873 (KS).

%
%
\bibliographystyle{apj}

\label{lastpage}

\end{document}